\documentclass[aps,prl,reprint,superscriptaddress,nofootinbib]{revtex4-2}

\usepackage{graphicx}
\usepackage{siunitx}
\usepackage{braket}
\usepackage{dcolumn}
\usepackage{bm}

\newcommand{\mytitlepage}{
    \begin{titlepage}
        \centering

        {\fontsize{12pt}{18pt}\selectfont\textbf{Studying Electron Beam Coherence Using Plasmon Interference}\par}
        {\fontsize{12pt}{18pt}\selectfont-- Supplementary Information --\par}
    \end{titlepage}
}

\begin{document}



\title{Studying Electron Beam Coherence Using Plasmon Interference}




\author{Evelijn Akerboom}
\affiliation{Center for Nanophotonics, NWO-Institute  AMOLF, 1098XG Amsterdam, The Netherlands}
\thanks{Correspondence to:\\
e.akerboom@amolf.nl\\
javier.garciadeabajo@nanophotonics.es\\
a.polman@amolf.nl\\}

\author{F. Javier Garc\'ia de Abajo}
\affiliation{ICFO-Institut de Ciencies
 Fotoniques, Barcelona, Spain and ICREA-Instituci\'o Catalana de Recerca i Estudis Avan\c{c}ats, Barcelona, Spain}
 
\author{Albert Polman}%
\affiliation{Center for Nanophotonics, NWO-Institute  AMOLF, 1098XG Amsterdam, The Netherlands}


\date{\today}

\begin{abstract}
Energetic electrons create distinct cathodoluminescence (CL) angular distributions upon interaction with dielectric and plasmonic nanostructures, providing valuable information on coherences in the excitation pathways. A counterintuitive prediction is that the CL signals arising from the excitation associated with different lateral regions of an extended electron wave are mutually incoherent and do not interfere, while the signals originating from different structures within the electromagnetic field of a narrow electron beam are mutually coherent. We present conclusive experimental evidence of these effects by examining the angular CL emission profile from defocused electron-beam excitation of a thin Si$_3$N$_4$ film, which is shown to follow an incoherent sum of CL excitations within the electron beam spot. In contrast, CL interferences are observed for separated plasmonic scatterers excited within the evanescent field of a single electron. Coherence may be recovered through correlations between emitted light and post-selected electron states, for which we propose a measurement geometry that erases which-path information of the electron trajectory.

\end{abstract}


\maketitle

Since the invention of the electron microscope \cite{ruska1987}, its applications have revolutionized many different fields, from imaging proteins \cite{Nogales2016,lorenz2024,henderson2015,Lorenz2013} and viruses \cite{adrian1984} to semiconductor metrology \cite{harvey2022,nasrazadani2016}. Due to their short de Broglie wavelength, swift free electrons provide a very small probe size, enabling atomic-scale imaging of crystal lattices and defects through advanced electron optics and detection techniques. In addition to their particle-like interactions, electrons exhibit wave properties, as first demonstrated by G. P. Thomson through electron diffraction from crystals \cite{thomson1931}. The wave nature enables techniques such as selected-area electron diffraction, electron backscatter diffraction, and convergent-beam electron diffraction (CBED), widely used to study crystal structures. \\
Besides yielding structural information, electron microscopes provide insight in optical material properties with nanometer spatial resolution through techniques such as electron energy-loss spectroscopy (EELS) and cathodoluminescence (CL). \cite{Nelayah2007,schaffer2009, Nicoletti2013, Coenen2013,mignuzzi2018energy,Matsukata2018,Sapienza2012}. EELS and CL data are successfully interpreted by regarding the electron as a classical point-charge particle that creates materials polarizations that lead to spectroscopic EELS and CL fingerprints. \\
The recently developed photon-induced near-field electron microscopy (PINEM) technique exploits the exchange of quanta between free electrons and strong optical near fields. Here, the electron acts as a quantum probe, and inelastic electron-light scattering shapes the electron wave packet in space and time \cite{Barwick2009}. This has stimulated several theoretical investigations \cite{garcia2010multi,Abajo2021,Giulio2019}, and recent work has experimentally demonstrated entanglement between the scattered electrons and the emitted CL photons \cite{henke2025eraser,bogdanov2025}, following theoretical predictions \cite{kazakevich2024,henke2025,frabboni,ruimy2024,konevcna2022}.\\
A key factor in the latter experiments is that the electron is described as a wave packet that can be shaped in transverse and longitudinal directions with respect to the electron beam (e-beam), in contrast to the classical point charge description typically used to explain EELS and CL measurements. The lateral and temporal coherence of the e-beam then determines the magnitude of the observed quantum effects. An open question is under which conditions CL excitation is sensitive to the coherence of the e-beam. Recent work has found that Smith–Purcell radiation from a grating is independent of transverse broadening of the e-beam \cite{remez2019}. However, a transversely coherent defocused e-beam may coherently excite optical scatterers that then interfere in the far field. The observation of such effects requires erasing of the quantum-mechanical which-path information of the multiple CL excitation paths. \\
In this Letter, we investigate how which-path information constrains observable interferences from coherent excitation of laterally separated electron paths. By studying the angular CL emission generated by both extended and localized electron excitations, we show that signals originating from different lateral regions of an extended electron wave do not interfere, yielding an incoherent superposition weighted by the electron density distribution. In contrast, excitation pathways involving spatially separated structures within the evanescent field of a single electron remain mutually coherent and give rise to observable CL interference. We first study transition radiation (TR) from a planar thin Si\textsubscript{3}N\textsubscript{4} film and characterize the angular CL profile for an extended e-beam spot size. While the e-beam has significant lateral coherence, the CL profiles represent an incoherent sum of TR across the e-beam spot. We then investigate a geometry in which a single electron couples to two plasmonic scatterers within the range of the electron's evanescent field, leading to the observation of coherent excitation of the two scatterers, as retrieved from clear far-field CL interferograms. Finally, we present a geometry in which the lateral coherence length is not sufficient to coherently excite two plasmonic scatterers and find a lack of interference, in agreement with theory. We end by proposing a geometry in which CL collection correlated with post-selected electron detection would erase the electron which-path information and lead to coherent CL excitation across the coherent part of the e-beam. \\

To describe the excitation by a laterally extended e-beam, we use a formalism valid when no post-selection of the electron is performed. In the nonrecoil approximation, the electron maintains its velocity and the interaction depends only on its transverse position. Under these conditions, the total excitation probability is obtained by summing over all possible final transverse electron states. This leads to a result where the excitation is an incoherent sum over different lateral positions of the electron, weighted by the transverse electron density (see Supplemental Material (SM) \cite{supplemental} for a self-contained theoretical description). As a consequence, no interference arises between different parts of the e-beam. However, if one were to select specific final electron states (e.g., a particular scattering direction), interference between different lateral components of the electron wave function can emerge. \\
\begin{figure}
	\centering
	\includegraphics{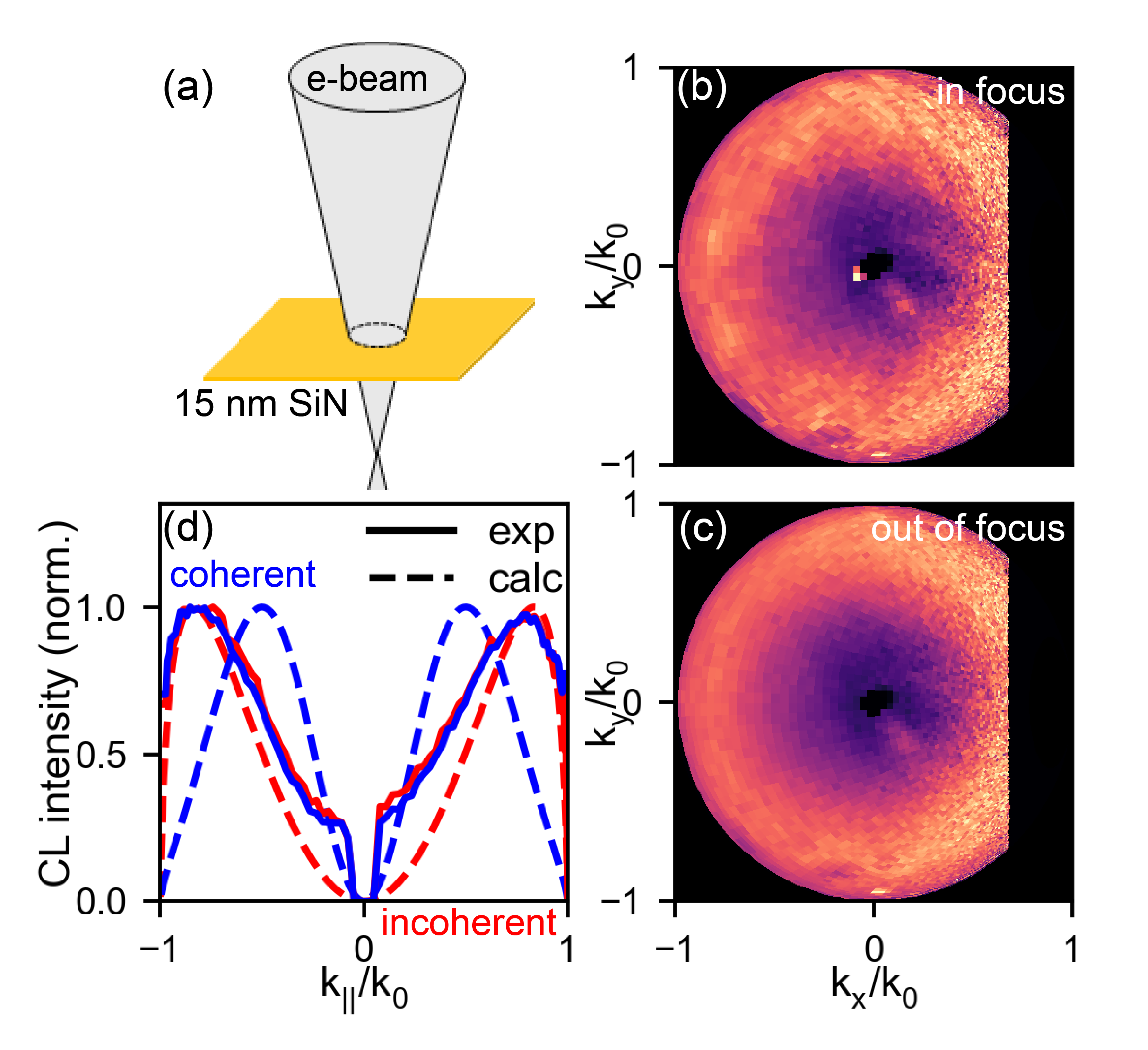}
	\caption{Lack of coherence in transition radiation excited by a laterally extended e-beam. We measure TR from a \SI{15}{}-\SI{}{\nano\meter} Si\textsubscript{3}N\textsubscript{4} thin film excited by \SI{30}{}-\SI{}{\kilo\electronvolt} electrons at a beam current of \SI{1.6}{\nano\ampere}. (a) Schematic representation of the experiment when the sample is placed above the focal point of the e-beam to laterally spread the electron spot at the sample. (b,c) AR CL data measured with a \SI{50}{}-\SI{}{\nano\meter} bandpass filter centered at a wavelength of \SI{500}{\nano \meter}, for (b) the Si\textsubscript{3}N\textsubscript{4} film in the focal position, and (c) with a spread e-beam of \SI{6}{}-\SI{}{\micro\meter} diameter. (d) Calculated (dashed) and experimentally measured (solid) data for TR emitted from the Si\textsubscript{3}N\textsubscript{4} film. The calculation shows both the characteristic AR CL emission for a point dipole (red), and the coherent superposition of TR summed for sources spread over a laterally coherent e-beam spot with a Gaussian distribution of \SI{150}{\nano\meter} standard deviation. The experimental curves show the almost indistinguishable angular distributions of TR emission from a focused e-beam (red) and the defocused e-beam (blue), which agree well with the incoherent sum. In panels (b-d), ${\textbf{k}}_\parallel=(k_x,k_y)$ is the in-plane wave vector of the emitted light, normalized to the wavenumber $k_0$.}
	\label{fig:Latcoh_fig1}
\end{figure}
We test whether lateral electron coherence modifies CL excitation through TR emitted from a thin Si\textsubscript{3}N\textsubscript{4} film. TR is a coherent emission process with a characteristic angle-dependent CL emission distribution (see SM\cite{supplemental} for an analytical description). To investigate how the lateral coherence of the e-beam affects TR emission, we measure the angle-resolved (AR) CL intensity under two conditions: (i) when the e-beam is tightly focused onto the sample, to a spot diameter of approximately \SI{4}{\nano\meter}, and (ii) when the sample is positioned above the e-beam’s focal plane, resulting in a defocused spot of approximately \SI{6}{}-\SI{}{\micro \meter} diameter. A schematic of the measurement geometries is shown in Fig. \ref{fig:Latcoh_fig1}(a). We use a \SI{30}{}-\SI{}{\kilo \electronvolt} e-beam and collect the AR CL emission through a \SI{50}{\nano\meter} bandpass filter centered at a wavelength of \SI{500}{\nano \meter}. The e-beam current is \SI{1.6}{\nano\ampere}, corresponding to an average arrival time between electrons at the sample of \SI{100}{\pico\second}, much larger than the TR decay time of a few \SI{}{\femto\second}. Figure \ref{fig:Latcoh_fig1}(d) shows the normalized angular emission calculated from Eq. (\ref{eq:TRanalytical}) for the focused beam (red-dashed curve) acting as a dipolar point source, and for the coherent sum of TR emitted over the laterally spread e-beam. The lateral coherence function of the used SEM is modeled through a Gaussian excitation distribution with a standard deviation of \SI{5}{\percent} of the full e-beam width, based on the measured degree of lateral coherence of the e-beam in our setup (see ref. \cite{akerboom2026} for a detailed analysis). As shown in Fig. \ref{fig:Latcoh_fig1}(d) for the coherent sum, the emission distribution (blue-dashed curve) shifts toward the surface normal compared to the characteristic TR angular profile excited by a point dipole. Clearly, the calculated coherent and incoherent TR emission show distinct angular distributions. \\
\begin{figure*}
	\includegraphics{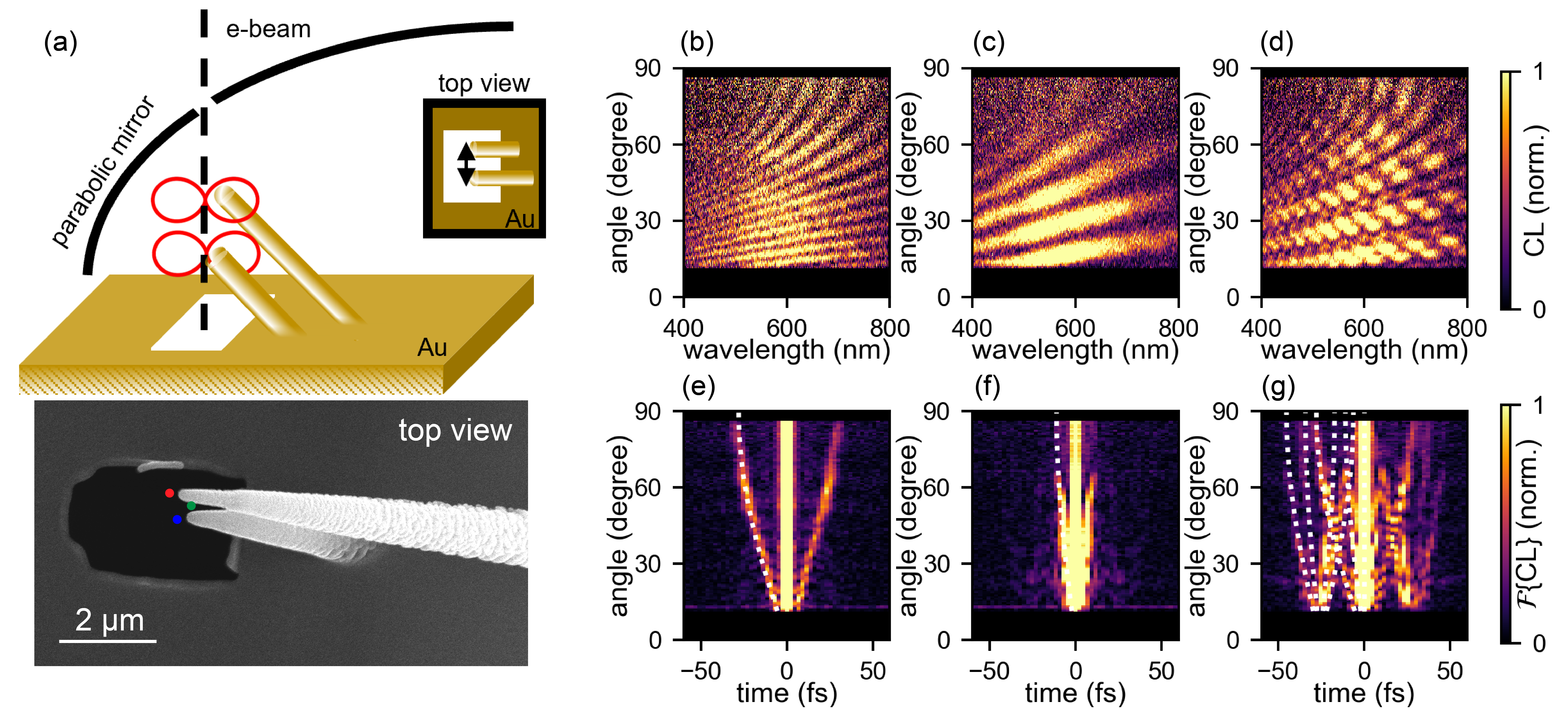}
	\caption{Narrow-beam excitation of two Au pillars with different lengths horizontally displaced by \SI{200}{\nano\meter}. (a) Schematic representation of the measurement (top) and top-view SEM image of the fabricated sample with the colored dots indicating the three measurement configurations (bottom). (b-d) AR spectral CL data from the double pillars and (e-g) Fourier-transformed data in the time domain for three different configurations: (b,e) the e-beam grazes the larger pillar (red dot in (a)), (c,f) the electron grazes the smaller pillar (blue dot in (a)), and (d,g) the e-beam passes in between the two pillars (green dot in (a)). The white dashed lines in (e-g) indicate the expected time delay components corresponding to pillar heights of \SI{4.2}{\micro \meter} and \SI{1.7}{\micro\meter} for the large and small pillars, respectively.}
	\label{fig:Latcoh_fig3}
\end{figure*}
Our experimental AR CL results for the focused and defocused beams are shown in Fig. \ref{fig:Latcoh_fig1}b,c, respectively, plotted as functions of the in-plane wave vector normalized to the free-space wavenumber $k_0=2\pi/\lambda$, with $\lambda = $\SI{500}{\nano \meter}. We observe similar angular emission distributions under both conditions. To better compare the angular emission profiles with the calculations, Fig. \ref{fig:Latcoh_fig1}(d) shows the measured intensity under in-focus (red) and out-of-focus (blue) conditions, integrated over an angular range of \SI{0.05}{\radian} along the $k_x/k_0=0$ direction. The similarity between the two measured curves indicates that the out-of-focus excitation represents an incoherent excitation of TR across the e-beam spot size [i.e., no lateral coherence in the excitation, as predicted by Eq. (\ref{eq:Latcoh_1a})]. The discrepancy between the theoretical and experimental data for the smaller wave vectors for the focused excitation is ascribed to contributions of scattered light from an impurity, visible in the $\textbf{k}_{\parallel}$-resolved images [Fig. \ref{fig:Latcoh_fig1}(b,c)] as a bright spot near the surface normal, which is superimposed on the coherent TR emission.  \\
These results show that, in a thin film, we measure the incoherent sum of TR emitted within the e-beam spot size, indicating that the transverse shape of the e-beam does not influence the angular distribution of the emitted CL, and lateral coherence in the incident e-beam does not translate into lateral coherence in the excitation. Despite the fundamental erasing of lateral coherence in the CL emission, here corroborated with theory (Eq. (\ref{eq:Latcoh_1a})) and experiment (Fig. \ref{fig:Latcoh_fig1}), the question arises: under which conditions is it possible to coherently excite multiple spatially displaced scatterers and observe interference in the far field?\\
\begin{figure*}
	\includegraphics{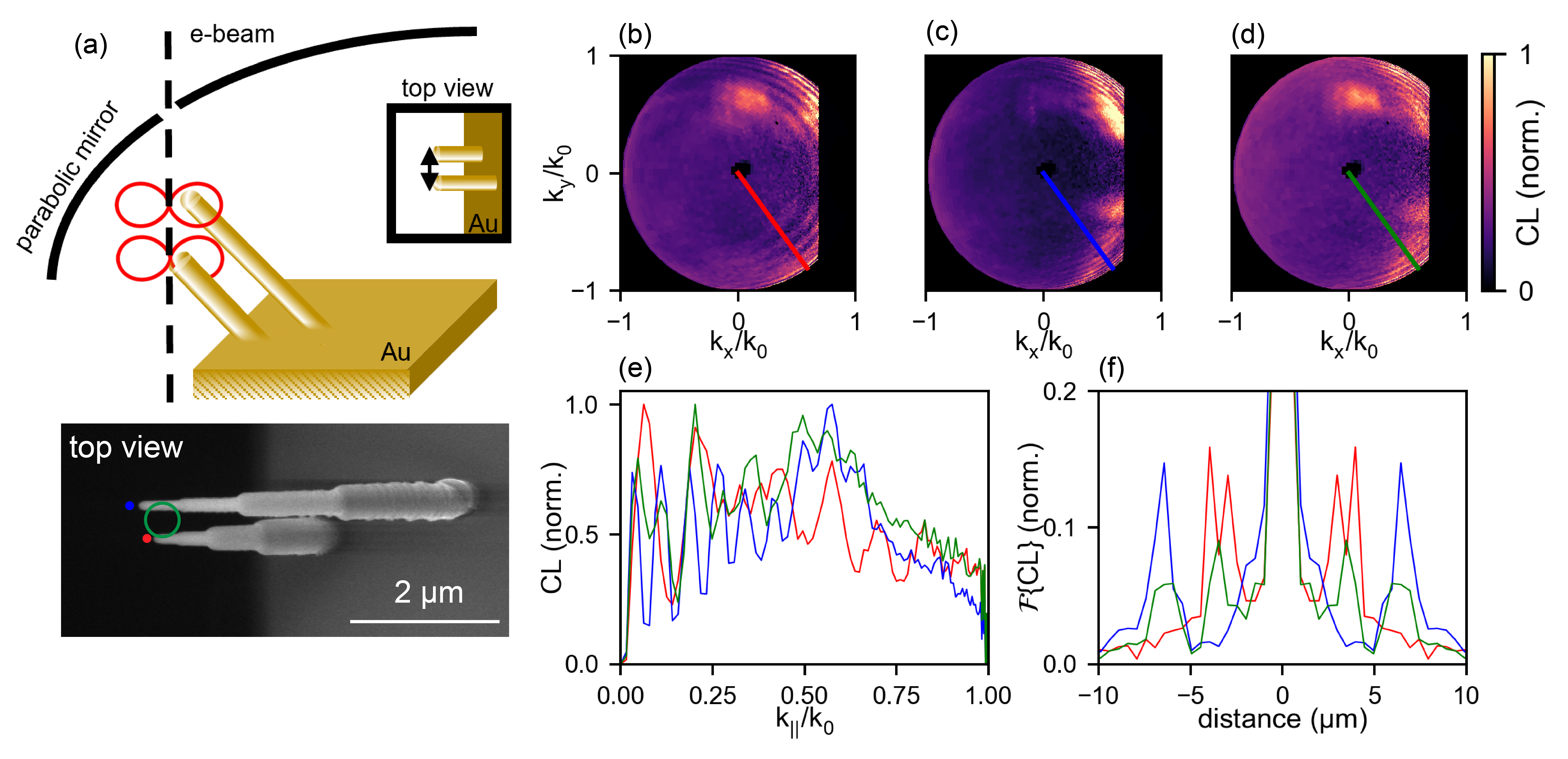}
	\caption{Excitation of a double pillar geometry with a laterally spread e-beam. (a) Schematic representation of the measurement (top) and top-view SEM image of the fabricated sample (bottom). (b-d) AR CL data from the double pillar geometry using a \SI{50}{}-\SI{}{\nano\meter} bandwidth filter centered around a wavelength of \SI{500}{\nano\meter}. (e) Line scans along the $\textbf{k}_{\parallel}$-space images and (f) their Fourier transform, taken from cuts along the solid lines in (b-d) for excitation by an e-beam passing near the smaller pillar (red) and the taller pillar (blue), or placed between the two pillars, with an e-beam diameter of \SI{500}{\nano\meter} (green). Curves in (e,f) are color-coordinated with the dots and circle in the SEM image of panel (a).}
	\label{fig:Latcoh_fig4}
\end{figure*}

To answer this question, we study a geometry consisting of two plasmonic nanopillars that are separated independently in the horizontal and vertical directions relative to the e-beam. We use vertical spacing to build up a phase difference of the CL emission and use a transversely spread e-beam with a diameter of a few hundred nanometers. As we have recently shown \cite{akerboom2025}, a single electron can coherently excite multiple scatterers that are spaced along the electron trajectory. Here, we create the plasmonic scattering tips by growing either a W or Pt nanopillar under a \SI{45}{\degree} tilt angle using electron beam induced deposition (EBID) in the SEM, and coat the sample with \SI{50}{\nano \meter} of Au (see SM \cite{supplemental} for fabrication details). \\
In a first scenario, we investigate a geometry in which we place the pillars next to each other, but purposely close enough for the evanescent field of the electron to couple to both of them when it passes in between the pillars. Furthermore, to eliminate the contribution of TR, we remove the substrate directly underneath the pillars using focused ion beam (FIB) milling. Figure \ref{fig:Latcoh_fig3}(a) shows a schematic representation of this geometry and a top-view SEM image of the resulting sample. The pillars, with different lengths, are horizontally separated by \SI{200}{\nano\meter}, resulting in a V-groove where we can precisely place the e-beam in the center and assess whether we can couple to both tips simultaneously. To study potential CL interference associated with simultaneous excitations at both pillars, we perform AR spectral CL measurements, where we measure the angular and spectral distribution of the CL emission simultaneously (see SM \cite{supplemental} for more details on the experiment). To analyze the components contributing to the observed interference, we take the Fourier transform at every emission angle, and thus, translate the data from the frequency domain to the time domain. \\
Figure \ref{fig:Latcoh_fig3}(b,c,and d) shows the measured experimental data for three respective configurations depending on the position of the e-beam focus relative to the two pillars: (1) near the larger pillar [red dot in (a)]; (2) near the smaller pillar [blue dot in (a)]; or (3) at a position in between the two pillars [green dot in (a)]. In Fig. \ref{fig:Latcoh_fig3}(b,c), for excitation of the large and small pillar, respectively, we observe interference patterns over the entire spectral range, which we assign to the superposition of the direct tip radiation and the reflected signal (i.e., the image of the excited tip produced by reflection on the substrate) \cite{akerboom2025}. By analyzing the Fourier transform of the data shown in Fig. \ref{fig:Latcoh_fig3}(e,f), we determine the height of the pillars to be \SI{4.2}{\micro\meter} and \SI{1.7}{\micro\meter}, respectively. \\
Figure \ref{fig:Latcoh_fig3}(d,g) shows data obtained when the e-beam passes between the two tips. Compared to the previous cases, we observe a richer interference pattern, with additional temporal features revealed in the Fourier transform. Besides the symmetric bands centered at $t=$\SI{0}{\femto \second}, we identify bands intersecting the time axis at $t=$\SI{25}{\femto \second}. This feature corresponds to the time of flight (TOF) of the electron traveling from the taller to the shorter pillar. The extracted TOF matches a height difference of \SI{2.5}{\micro\meter}, in agreement with the known geometry of the sample. The presence of this TOF component indicates coherent excitation of both pillars when the electron passes between them. In this configuration, a single electron simultaneously excites both structures when they are both within the range of the electron's evanescent field. Importantly, this coherence does not arise from lateral coherence of the electron wave packet itself, but from simultaneous coupling of the electron near field to both pillars, consistent with a classical point-particle description of the electron. Due to the small height difference, the interference bands overlap in the Fourier transform, and it is difficult to identify their individual contributions. We have performed a similar analysis on a system with larger structures to fully decouple the different components (see Supplemental Material \cite{supplemental}).

Next, we place the pillars further apart than twice the extent of the electron's evanescent field and defocus the e-beam such that its spot overlaps with both pillars. To eliminate the interference resulting from either emitted TR or reflection at the substrate, we partly remove the Au-coated membrane using FIB milling. As a result, the only remaining interference must originate from coherent excitation of the two tips.\\Figure \ref{fig:Latcoh_fig4}(a) shows a schematic representation of the sample. The nanopillars are spaced \SI{400}{\nano\meter} in the horizontal direction, as indicated by the top view. The SEM image shows the fabricated sample, where, on the left side, the substrate is removed using FIB milling. To study the coherent or incoherent nature of the excitation, we carry out AR CL measurements using a \SI{50}{}-\SI{}{\nano\meter} bandwidth filter centered at \SI{500}{\nano\meter}. We perform the measurement in three configurations: (1) when the focused e-beam grazes the small pillar (red dot), or (2) the tall pillar (blue dot), and (3) when the electron passes through the middle of the two pillars and is laterally extended to a diameter of approximately \SI{500}{\nano\meter} (green circle). \\
Figure \ref{fig:Latcoh_fig4}(b) and \ref{fig:Latcoh_fig4}(c) show the measured AR CL data for an electron grazing the small and large pillars, respectively. Interference fringes are observed for positive $k_x$, corresponding to the right side of the SEM image in Fig. \ref{fig:Latcoh_fig4}(a), where the substrate is present. These fringes arise from interference between direct tip emission and its reflection at the substrate, similar to the data in Fig. \ref{fig:Latcoh_fig3}. To determine the pillar height, we take a line scan of the $\textbf{k}_{\parallel}$-space, shown in Fig. \ref{fig:Latcoh_fig4}(e) for the small (red) and large (blue) pillar. We clearly see the interferences at low emission angles and take the Fourier transform to translate this to the spatial domain, yielding pillar heights of \SI{1.7}{} and \SI{3.3}{\micro \meter}.\\
Next, the e-beam is positioned between the tips and defocused to a spot size of approximately \SI{500}{\nano\meter}, covering both tips simultaneously. The coherent area of the beam spans around \SI{25}{\nano\meter} (\SI{5}{\percent} of the full e-beam width, based on the measured degree of lateral coherence of the e-beam in our setup)  \cite{akerboom2026}, and therefore, we do not expect to see interference from two coherently excited pillars. The AR CL data [Fig. \ref{fig:Latcoh_fig4}(d)] show clear fringes in the same angular range where they were observed in Fig. \ref{fig:Latcoh_fig4}(b,c). When we take the Fourier transform of a line scan of this data, as shown in Fig. \ref{fig:Latcoh_fig4}(e,f), respectively (green), we clearly observe peaks associated with both the taller and the smaller pillars, indicating that the e-beam excites both of them. However, the absence of interference at negative $k_x$ values, on the side where the substrate is removed, indicates that the CL signal recorded in the angular range represents an incoherent sum of the radiation from both pillars (see Supplemental Material for more details \cite{supplemental}). This is consistent with both the limited lateral coherence of the e-beam and the fact that the which-path information in the transmitted electrons is not erased in this geometry. \\

These results show that, while coherent excitation of plasmonic scatterers can be observed for a single electron exciting multiple scatterers within its evanescent field, coherent TR excitation is not observed for an expanded e-beam even when it has sufficient lateral coherence to observe the effect. Observing coherent excitation requires erasing the electron which-path information. In the geometry studying TR from an expanded e-beam spot, the electron effectively \textit{knows} the position at which TR is excited, as this information could, in principle, be measured by determining the position of the electron within the e-beam once its wave function has collapsed by creating localized excitations. Recent experiments in which which-path information is erased in electron-photon scattering and correlation are reported by Henke \textit{et al.} \cite{henke2025eraser}. There, polarization information contained within scattered photons was erased by mixing polarized signals in order to retrieve the interference on the electron detection plane. In our case, the situation is inverted: we must erase the information carried by the electrons in order to retrieve interference of the CL photons. A practical way to achieve this would be to place a structure underneath the double pillar that diffracts the electrons, such as a twisted bilayer of graphene (see Supplemental Material for a schematic of the proposed experiment \cite{supplemental}). The diffraction pattern from this structure would consist of partially overlapping diffraction disks. In those overlapping regions, the electron could originate from multiple places within the e-beam, and therefore, the spatial information is erased. By post-selecting electrons in the overlapping diffraction region and correlating them with detected photons, we expect to retrieve interference of coherently excited CL signals across the coherent area of the e-beam spot. Implementing this scheme requires dedicated sample fabrication and electron–photon correlation capabilities, both of which are experimentally challenging, but feasible.

\begin{acknowledgments}
\textit{Acknowledgments---}We thank Nick J. Schilder for initial AR-TR experiments. This work is financed by the Dutch Research Council (NWO) and has received funding from the European Research Council (ERC) under Grant Agreements Nos. 101019932 (QEWS) and 101141220 (QUEFES), and F.J.G.A. acknowledges support from the Spanish MICIU (PID2024-157421NB-I00 and Severo Ochoa CEX2024-001490-S), and the Catalan CERCA Program.
\end{acknowledgments}

\bibliography{apssamp}

@article{schaffer2009,
  title={High-resolution surface plasmon imaging of gold nanoparticles by energy-filtered transmission electron microscopy},
  author={Schaffer, Bernhard and Hohenester, Ulrich and Tr{\"u}gler, Andreas and Hofer, Ferdinand},
  journal={Physical Review B—Condensed Matter and Materials Physics},
  volume={79},
  number={4},
  pages={041401},
  year={2009},
  publisher={APS}
}

@article{mignuzzi2018energy,
  title={Energy--momentum cathodoluminescence spectroscopy of dielectric nanostructures},
  author={Mignuzzi, Sandro and Mota, Monica and Coenen, Toon and Li, Yi and Mihai, Andrei P and Petrov, Peter K and Oulton, Rupert FM and Maier, Stefan A and Sapienza, Riccardo},
  journal={ACS Photonics},
  volume={5},
  number={4},
  pages={1381--1387},
  year={2018},
  publisher={ACS Publications}
}

@article{garcia2010multi,
  title={Multiphoton absorption and emission by interaction of swift electrons with evanescent light fields},
  author={Garc{\'\i}a de Abajo, F Javier and Asenjo-Garcia, Ana and Kociak, Mathieu},
  journal={Nano Letters},
  volume={10},
  number={5},
  pages={1859--1863},
  year={2010},
  publisher={ACS Publications}
}

@article{Abajo2021,
   author = {Garc\'ia de Abajo, F Javier and Di Giulio, Valerio},
   doi = {10.1021/acsphotonics.0c01950},
   issn = {23304022},
   issue = {4},
   journal = {ACS Photonics},
   month = {8},
   pages = {945-974},
   publisher = {American Chemical Society},
   title = {Optical Excitations with Electron Beams: Challenges and Opportunities},
   volume = {8},
   year = {2021}
}

@article{konevcna2022,
  title={Entangling free electrons and optical excitations},
  author={Kone{\v{c}}n{\'a}, Andrea and Iyikanat, Fadil and Garc{\'\i}a de Abajo, F Javier},
  journal={Science Advances},
  volume={8},
  number={47},
  pages={eabo7853},
  year={2022},
  publisher={American Association for the Advancement of Science}
}

@article{Barwick2009,
   author = {Brett Barwick and David J Flannigan and Ahmed H Zewail},
   doi = {10.1038/nature08662},
   issn = {00280836},
   issue = {7275},
   journal = {Nature},
   pages = {902-906},
   pmid = {20016598},
   publisher = {Nature Publishing Group},
   title = {Photon-induced near-field electron microscopy},
   volume = {462},
   year = {2009}
}

@article{Coenen2013,
  title={Resonant modes of single silicon nanocavities excited by electron irradiation},
  author={Coenen, Toon and van de Groep, Jorik and Polman, Albert},
  journal={ACS nano},
  volume={7},
  number={2},
  pages={1689--1698},
  year={2013},
  publisher={ACS Publications}
}

@article{Giulio2019,
   author = {Di Giulio, Valerio and Kociak, Mathieu and Garc\'ia de Abajo, F Javier},
   doi = {10.1364/optica.6.001524},
   issn = {23342536},
   issue = {12},
   journal = {Optica},
   pages = {1524},
   title = {Probing quantum optical excitations with fast electrons},
   volume = {6},
   year = {2019}
}

@article{Sapienza2012,
   author={Sapienza, Riccardo and Coenen, Toon and Renger, Jan and Kuttge, Martin and van Hulst, Niek F and Polman, A},
   doi = {10.1038/nmat3402},
   issn = {14764660},
   issue = {9},
   journal = {Nature Materials},
   pages = {781-787},
   publisher = {Nature Publishing Group},
   title = {Deep-subwavelength imaging of the modal dispersion of light},
   volume = {11},
   year = {2012}
}

@article{Matsukata2018,
   author = {Taeko Matsukata and Carl Wadell and Nikolaos Matthaiakakis and Naoki Yamamoto and Takumi Sannomiya},
   doi = {10.1021/acsphotonics.8b01231},
   issn = {23304022},
   issue = {12},
   journal = {ACS Photonics},
   pages = {4986-4992},
   title = {Selected Mode Mixing and Interference Visualized within a Single Optical Nanoantenna},
   volume = {5},
   year = {2018}
}

@article{Nelayah2007,
   author = {Jaysen Nelayah and Mathieu Kociak and Odile Stéphan and F Javier {García de Abajo} and Marcel Tencé and Luc Henrard and Dario Taverna and Isabel Pastoriza-Santos and Luis M Liz-Marzán and Christian Colliex},
   doi = {10.1038/nphys575},
   issn = {17452481},
   issue = {5},
   journal = {Nature Physics},
   pages = {348-353},
   publisher = {Nature Publishing Group},
   title = {Mapping surface plasmons on a single metallic nanoparticle},
   volume = {3},
   year = {2007}
}

@article{Nicoletti2013,
  title={Three-dimensional imaging of localized surface plasmon resonances of metal nanoparticles},
  author={Nicoletti, Olivia and de La Pe{\~n}a, Francisco and Leary, Rowan K and Holland, Daniel J and Ducati, Caterina and Midgley, Paul A},
  journal={Nature},
  volume={502},
  number={7469},
  pages={80--84},
  year={2013},
  publisher={Nature Publishing Group UK London}
}

@article{GarcaDeAbajo2010,
  title={Optical excitations in electron microscopy},
  author={Garc{\'\i}a de Abajo, Francisco Javier},
  journal={Reviews of Modern Physics},
  volume={82},
  number={1},
  pages={209--275},
  year={2010},
  publisher={APS}
}

@article{ruska1987,
  title={The development of the electron microscope and of electron microscopy},
  author={Ruska, Ernst},
  journal={Reviews of Modern Physics},
  volume={59},
  number={3},
  pages={627},
  year={1987},
  publisher={APS}
}

@article{nogales2016,
  title={The development of cryo-{EM} into a mainstream structural biology technique},
  author={Nogales, Eva},
  journal={Nature Methods},
  volume={13},
  number={1},
  pages={24--27},
  year={2016},
  publisher={Nature Publishing Group US New York}
}

@article{lorenz2024,
  title={Microsecond time-resolved cryo-electron microscopy},
  author={Lorenz, Ulrich J},
  journal={Current Opinion in Structural Biology},
  volume={87},
  pages={102840},
  year={2024},
  publisher={Elsevier}
}

@article{henderson2015,
  title={Overview and future of single particle electron cryomicroscopy},
  author={Henderson, Richard},
  journal={Archives of biochemistry and biophysics},
  volume={581},
  pages={19--24},
  year={2015},
  publisher={Elsevier}
}

@article{adrian1984,
  title={Cryo-electron microscopy of viruses},
  author={Adrian, Marc and Dubochet, Jacques and Lepault, Jean and McDowall, Alasdair W},
  journal={Nature},
  volume={308},
  number={5954},
  pages={32--36},
  year={1984},
  publisher={Nature Publishing Group UK London}
}

@article{harvey2022,
  title={Using Benchtop Scanning Electron Microscopy as a Valuable Imaging Tool in Various Applications},
  author={Harvey, Kayleigh and Edwards, Grant},
  journal={Microscopy Today},
  volume={30},
  number={5},
  pages={32--35},
  year={2022},
  publisher={Oxford University Press}
}

@incollection{nasrazadani2016,
title = {Chapter 2 - Modern analytical techniques in failure analysis of aerospace, chemical, and oil and gas industries},
editor = {Abdel Salam Hamdy Makhlouf and Mahmood Aliofkhazraei},
booktitle = {Handbook of Materials Failure Analysis with Case Studies from the Oil and Gas Industry},
publisher = {Butterworth-Heinemann},
pages = {39-54},
year = {2016},
isbn = {978-0-08-100117-2},
doi = {https://doi.org/10.1016/B978-0-08-100117-2.00010-8},
url = {https://www.sciencedirect.com/science/article/pii/B9780081001172000108},
author = {Seifollah Nasrazadani and Shokrollah Hassani},
}

@article{thomson1931,
  title={The diffraction of electrons by single crystals},
  author={Thomson, George Paget},
  journal={Proceedings of the Royal Society of London. Series A, Containing Papers of a Mathematical and Physical Character},
  volume={133},
  number={821},
  pages={1--25},
  year={1931},
  publisher={The Royal Society London}
}

@article{kazakevich2024,
  title={Spatial electron-photon entanglement},
  author={Kazakevich, Eitan and Aharon, Hadar and Kfir, Ofer},
  journal={Physical Review Research},
  volume={6},
  number={4},
  pages={043033},
  year={2024},
  publisher={APS}
}

@article{henke2025,
  title={Probing electron-photon entanglement using a quantum eraser},
  author={Henke, Jan-Wilke and Jeng, Hao and Ropers, Claus},
  journal={Physical Review A},
  volume={111},
  number={1},
  pages={012610},
  year={2025},
  publisher={APS}
}

@article{frabboni,
  title={Ion and electron beam nanofabrication of the which-way double-slit experiment in a transmission electron microscope},
  author={Frabboni, Stefano and Gazzadi, Gian Carlo and Pozzi, Giulio},
  journal={Applied Physics Letters},
  volume={97},
  number={26},
  pages={263101},
  year={2010},
  publisher={AIP Publishing}
}

@article{ruimy2024,
  title={Many-body entanglement via ‘which-path’ information},
  author={Ruimy, Ron and Tziperman, Offek and Gorlach, Alexey and M{\o}lmer, Klaus and Kaminer, Ido},
  journal={npj Quantum Information},
  volume={10},
  number={1},
  pages={121},
  year={2024},
  publisher={Nature Publishing Group UK London}
}

@article{henke2025eraser,
  title={Observation of quantum entanglement between free electrons and photons},
  author={Henke, Jan-Wilke and Jeng, Hao and Sivis, Murat and Ropers, Claus},
  journal={arXiv:2504.13047},
  year={2025}
}

@misc{akerboom2026,
      title={Determining Electron Beam Lateral Coherence in a Scanning Electron Microscope Using Electron Diffraction}, 
      author={Akerboom, Evelijn and Kiani, Fatemeh and Tagliabue, Giulia and Albrecht, Wiebke and Etheridge, Joanne and García de Abajo, F. Javier and Polman, Albert},
      year={2026},
      eprint={2606.28056},
      archivePrefix={arXiv},
      primaryClass={cond-mat.mes-hall},
      url={https://arxiv.org/abs/2606.28056}, 
}

@article{bogdanov2025,
  title={Ghost Imaging with Free Electron-Photon Pairs},
  author={Bogdanov, Sergei and Preimesberger, Alexander and Mishra, Harsh and Hornof, Dominik and Spielauer, Thomas and Thajer, Florian and Maurer, Max and Falb, Pia and St{\"o}ger, Leo and Schachinger, Thomas and Bleicher, Friedrich and Seifner, Michael S. and Bicket, Isobel C. and Haslinger, Philipp},
  journal={arXiv:2509.14950},
  year={2025}
}

@article{remez2019,
  title={Observing the quantum wave nature of free electrons through spontaneous emission},
  author={Remez, Roei and Karnieli, Aviv and Trajtenberg-Mills, Sivan and Shapira, Niv and Kaminer, Ido and Lereah, Yossi and Arie, Ady},
  journal={Physical review letters},
  volume={123},
  number={6},
  pages={060401},
  year={2019},
  publisher={APS}
}

@article{akerboom2025,
  title={Angle-Resolved Cathodoluminescence Interferometry of Plasmonic and Dielectric Scatterers},
  author={Akerboom, Evelijn and Sugimoto, Hiroshi and Fujii, Minoru and Garc\'ia de Abajo, F Javier and Polman, Albert},
  journal={Nano Letters},
  volume={25},
  number={39},
  pages={14264--14269},
  year={2025},
  publisher={ACS Publications}
}

@article{Lorenz2013,
author = {Lorenz, Ulrich J. and Zewail, Ahmed H.},
doi = {10.1073/pnas.1300630110},
issn = {00278424},
journal = {Proceedings of the National Academy of Sciences of the United States of America},
number = {8},
pages = {2822--2827},
pmid = {23382239},
title = {{Biomechanics of DNA structures visualized by 4D electron microscopy}},
volume = {110},
year = {2013}
}

@misc{supplemental,
title = {See Supplemental Material},
howpublished = {},
note = {for a self-contained derivation of the theoretical formalism, as well as further details on sample fabrication and experimental measurements.}
}

\newpage 

\appendix

\onecolumngrid

\section*{End Matter}


\twocolumngrid

\section{Lack of lateral coherent electron excitation: Theoretical insight}
\label{app:sec2}
To theoretically describe the excitation of a laterally extended e-beam we follow a formalism that is generally valid when no post-selection of the electron is performed \cite{Abajo2021}. We study the interaction between an electron and a structure that supports optical excitations of low energy compared to the electron energy. Therefore, we can assume that the electron does not change its velocity along the e-beam trajectory (nonrecoil approximation). Under these conditions, the effective interaction Hamiltonian $\hat{H}(\textbf{R})$ only depends on the transverse coordinate $\mathbf{R}$. Since the transverse electron energy transfer is negligible under the non-recoil approximation, the total excitation probability ($P$) is the incoherent sum over all final states with transverse wave vectors $\textbf{Q}$ (note that the longitudinal wave vector is determined by energy conservation \cite{GarcaDeAbajo2010}), given by 
\begin{equation}
P\propto \int d^2\textbf{Q}\left|\int d^2 \textbf{R} e^{-i\textbf{Q}\cdot\textbf{R}}\braket{n|\hat{H}(\textbf{R})|0}\psi_{\perp}^i(\textbf{R})\right|^2. 
\end{equation}
 Here, $\mathbf{Q}$ is the 2D wave vector of the electron after interaction (i.e., the component in the plane perpendicular to the e-beam), while $\psi_{\perp}^i$ is the incident transverse wave function of the electron. Since the matrix element is a Fourier transform (from $\textbf{R}$ to $\textbf{Q}$), integrating over $\textbf{Q}$ yields 
\begin{equation}
    P \propto \int d^2\textbf{R} \left| \braket{n|\hat{H}(\textbf{R})|0}\right|^2 \left| \psi_{\perp}^i(\textbf{R})\right|^2
    \label{eq:Latcoh_1a}
\end{equation}
according to the Plancherel theorem. This demonstrates that the total probability does not contain any interference between different lateral positions of the electron (i.e., it is the incoherent superposition of different lateral positions, weighted by the transverse electron density profile). This conclusion would change if we were to measure transitions associated with a single final state (i.e., a fixed $\textbf{Q}$, or equivalently, an outgoing electron scattering direction) or a small range of wave vectors $\textbf{Q}$. In that case, Eq. (\ref{eq:Latcoh_1a}) no longer applies, and interference terms can appear.

\section{Theoretical description TR}
\label{app:sec1}
TR is a coherent form of CL emission, with a characteristic angular emission distribution. The electric field can be calculated as function of far-field position ($\textbf{r}=(x,y,z)$ and optical frequency ($\omega$) as \cite{GarcaDeAbajo2010} 
\begin{equation}
    \textbf{E}_\text{TR}(\textbf{r},\omega)=\textbf{f}(\textbf{r},\omega)\frac{e^{ik_0r}}{r},
\end{equation}
where $r=|\textbf{r}|$ and $\textbf{f}(\textbf{r},\omega)$ is given by 
\begin{equation}
    \textbf{f}(\textbf{r},\omega)=ik_0\cos(\theta)D\mu_1 \hat{\theta}
\end{equation}
where $k_0$ is the vacuum light wavenumber ($k_0=2\pi/\lambda$) and $D$ is given by 
\begin{equation}
 D = \frac{2ieq_{||}/c}{q_{z1}\epsilon_2 + q_{z2}\epsilon_1}. 
\end{equation}
Here, $q_{||}$ is the in-plane wave vector of light in vacuum, while $\mu_1$ is given by
\begin{equation}
 \mu_1 = \frac{1}{|v|}\left(\frac{\omega \epsilon_2 + v q_{z2}\epsilon_1}{q^2-k^2\epsilon_1} - \frac{\omega \epsilon_1 + v q_{z1}\epsilon_2}{q^2-k^2\epsilon_2}\right).
\end{equation}
When the e-beam is transversely broadened, we integrate the emission the e-beam spot size with radius ($R_e$), using a Gaussian distribution with a standard deviation ($\sigma$). The resulting electric field in the far field ($r>>R_e$) is given by 
\begin{equation}
    E_\text{TR}(\textbf{r},\omega)= f(\textbf{r},\omega)\frac{e^{ik_0r}}{r}\int d^2\mathbf{R_e} e^{ik_0\mathbf{r}\cdot \mathbf{R_e}}e^{-\frac{1}{2}R_e^2/\sigma^2}. 
\end{equation}
Solving the integral and taking the absolute square we find the CL emission probability per solid angle per electron to be proportional to 
\begin{equation}
    \Gamma_\text{CL} \propto |\textbf{f}(\textbf{r},\omega)|^2 e^{(k_0^\parallel/\sigma)^2},
    \label{eq:TRanalytical}
\end{equation}
where $k_0^\parallel$ is the parallel component of the free-space wavenumber ($k_0^\parallel = k_0\sin(\theta)$). To calculate the TR emission for a laterally spread e-beam in Fig. \ref{fig:Latcoh_fig1}, we use a standard deviation of \SI{150}{\nano\meter} and integrate the wavelength from \SI{475}{\nano\meter} to \SI{525}{\nano\meter}.

\section{Experimental setup and methods}
\label{sec:experimental}

\subsection{Angle- and spectrally resolved CL measurements }
\label{ch:Cohlat_chapter_ARspecCL}
The AR spectral CL measurements were performed in an FEI Quanta FEG 650 scanning electron microscope (SEM; Thermo Fisher Scientific Inc., MA, USA) equipped with a Schottky electron source using electrons with an energy of \SI{30}{\kilo\electronvolt} and a current of \SI{1.5}{\nano\ampere}. To perform the AR spectral CL measurements, the same collection system was used as for the AR CL measurements using a vertical slit to select only a narrow radial angle. Next, this was projected onto a spectrometer to acquire the CL data depending on wavelength and azimuthal angle simultaneously. For the double pillars (Fig. \ref{fig:Latcoh_fig3}), an exposure time of \SI{300}{\second} was used. In all cases the substrate was in the focal point of the parabolic mirror.

\subsection{Angle-resolved CL measurements}
\label{ch:Cohlat_chapter_ARCL}
The AR CL measurements are performed in a Helios5UX scanning electron microscope (SEM; Thermo Fisher Scientific Inc., MA, USA), equipped with a Schottky FEG source. The electron energy used was \SI{30}{\kilo \electronvolt}, with a beam current of \SI{1.6}{\nano\ampere}. For the AR CL measurements, a half-parabolic mirror was placed in between the sample and the pole piece, which collected the emitted light and directed it onto a camera (SPARC Spectral, DELMIC BV, The Netherlands). From the known parabolic geometry of the mirror, the image on the camera can be converted to the angular emission pattern. For the AR CL measurements a \SI{50}{}-\SI{}{\nano\meter} bandwidth filter centered at a wavelength of \SI{500}{\nano\meter} was used, and an exposure time of \SI{300}{\second} per pixel was used. For the TR from a thin film of Si\textsubscript{3}N\textsubscript{4} film, the e-beam was first focused on the surface, and for the defocused measurement, the focal point of the e-beam was placed \SI{2}{\milli\meter} below the sample. We used SEM images of the sharp edge of the Si\textsubscript{3}N\textsubscript{4} to estimate that the e-beam spot diameter was \SI{6}{\micro\meter}. \\
For the double pillar configuration shown in Fig. \ref{fig:Latcoh_fig4}, the e-beam current was \SI{0.8}{\nano\ampere}, an electron energy of \SI{30}{\kilo\electronvolt}, and an exposure time of \SI{300}{\second}. The out-of-focus measurement was done with the sample \SI{84}{\micro\meter} above the focal point, corresponding to an estimated e-beam spot diameter of \SI{600}{\nano\meter}. 

\subsection{Sample fabrication}
\label{ch:Cohlat_chapter_EBID}
The samples were fabricated using electron beam induced deposition (EBID) in a Helios5UX scanning electron microscope (SEM; Thermo Fisher Scientific Inc., MA, USA). The used substrate was a piece of single-crystalline Si for the double pillars on top of each other and a \SI{15}{}-\SI{}{\nano\meter}-thick Si\textsubscript{3}N\textsubscript{4} TEM grid (Ted Pella), both covered with \SI{50}{\nano\meter} of Au using a \SI{2}{}-\SI{}{\nano\meter} thin adhesion layer of Cr (EM ACE600, Leica, Inc.). A Pt or W pillar was grown under a \SI{45}{\degree} tilt angle, using a \SI{2}{}-\SI{}{\kilo\electronvolt} e-beam. Placing the pillars next to each other and increasing the dwell time for one pillar resulted in displaced pillars in the horizontal direction, with different heights. Next, in case we wanted to eliminate the contribution from TR (Fig. \ref{fig:Latcoh_fig3} and Fig. \ref{fig:Latcoh_fig4}), focused ion beam milling (FIB) was used to remove the Si\textsubscript{3}N\textsubscript{4} window underneath, after which the entire sample was coated with \SI{50}{\nano\meter} of Au.

\clearpage
\mytitlepage 
\onecolumngrid
\section{Coherent excitation of double pillars along the electron beam}
\begin{figure*}[!b]
	\centering
	\includegraphics{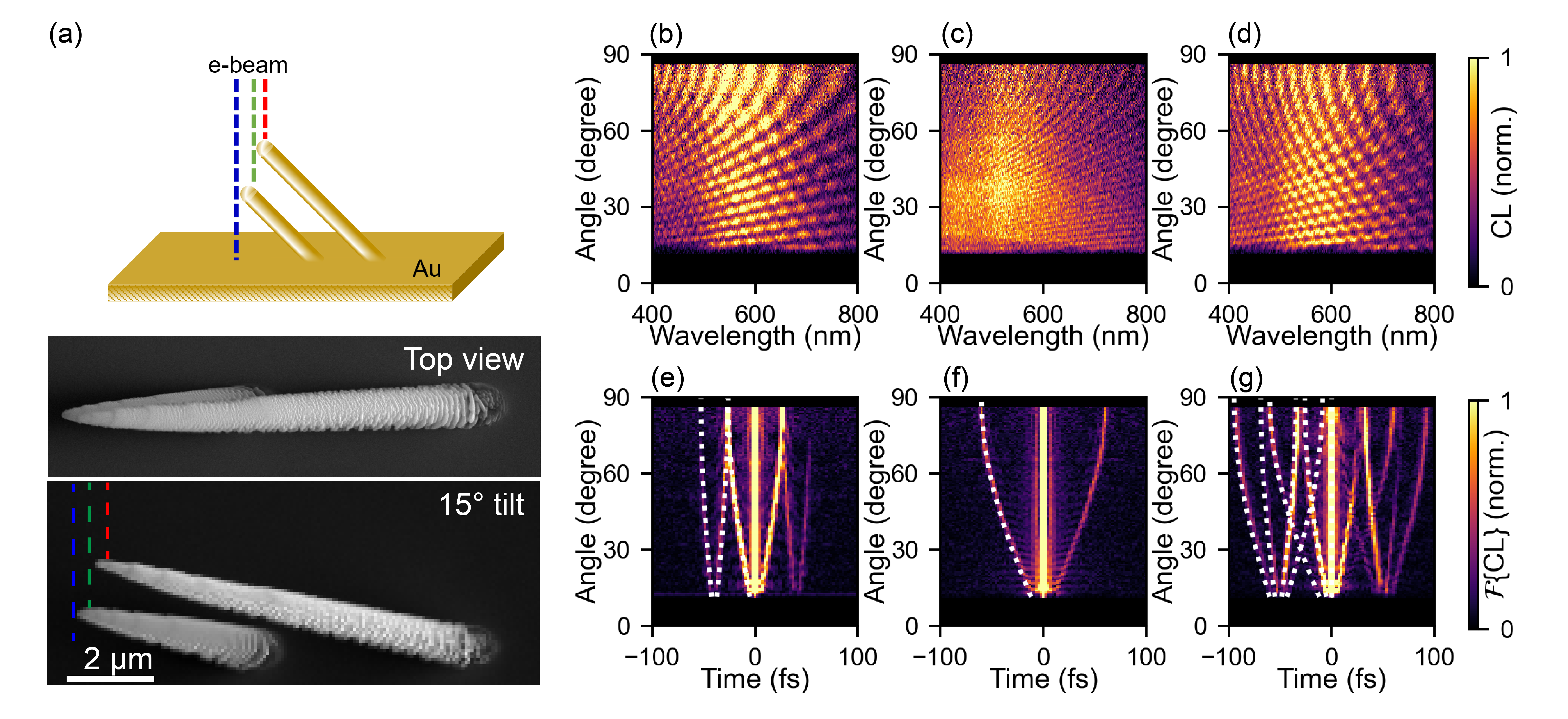}
	\caption{Coherent excitation of two Au pillars placed above each other. (a) Schematic representation of the measurement and the SE image of the fabricated sample taken from the top and at a tilt angle of \SI{15}{\degree}. (b-d) AR spectral CL data from the double pillars and (e-g) their Fourier transformed data in the time domain for three configurations: (b,e) the e-beam grazes the small pillar, (c,f) the electron impinges on the large pillar, and (d,g) the e-beam grazes the tall pillar and impinges on the smaller one. The white dashed lines in (e-g) indicate the calculated time delay components corresponding to a pillar height of \SI{3.9}{\micro \meter} and \SI{9.0}{\micro \meter} for the small and large pillar, respectively. }
	\label{fig:Latcoh_fig2}
\end{figure*}

To decouple the components in an interferogram of two plasmonic nanotips excited by one electron, we have made a sample where two nanotips are placed on top of each other with an Au substrate underneath. We perform AR spectral CL measurements to study the CL interference. To analyze the different components that are contributing to the observed interference, we take the Fourier transform at every emission angle, to translate the data from the frequency domain to the time domain.  \\
Figure \ref{fig:Latcoh_fig2} shows the experimental data for three configurations: (1) where the electron beam (e-beam) grazes the small pillar (blue), (2) when placed on top of the larger pillar (red), and (3) for an electron grazing the larger pillar, and impinging on the small pillar (green). Figure \ref{fig:Latcoh_fig2}(b,e) show the experimental AR spectral data and their Fourier transforms for the first scenario where the electron grazes the small pillar. In the spectral data, we clearly observe a cross-interference pattern indicating that more than two sources of radiation are excited by the e-beam. Examining the Fourier transform, we identify these components, in order from small to large time delay as (1) the interference between the direct tip radiation and the signal reflected from the surface, (2) the interference of TR emitted from the Au surface with tip radiation reflected from the substrate, and (3) the interference of TR emission with the direct tip radiation. From this interference pattern, we find a pillar height of \SI{3.9}{\micro \meter}.\\
Next, Figs. \ref{fig:Latcoh_fig2}(c) and \ref{fig:Latcoh_fig2}(f) show the AR spectral data and their Fourier transform, respectively, for the second scenario in which the electrons impinge on the large pillar. Instead of a cross-shaped pattern, we only find a single curved interference pattern, indicating that we only measure the interference of tip radiation and its reflection from the Au substrate \cite{akerboom2025}. The Fourier transform further corroborates this conclusion, where we find only one contribution, which we match to a pillar height of \SI{9.0}{\micro \meter}.\\
Next, we analyze the configuration of the electron grazing the tall pillar and impinging on the smaller one, shown in Fig. \ref{fig:Latcoh_fig2}(d,g). Here, we observe a richer and more complex interference pattern compared to the previous data, further supported by several additional components in the Fourier-transformed data. We identify two components that originate at $t=0$ for \SI{0}{\degree} emission angle that correspond to the interference features identified above, arising from the direct and reflected tip radiation from both pillars. In addition, we observe four bands originating at $t=$\SI{51}{\femto\second} at \SI{0}{\degree} emission angle. These represent the interference of radiation from the two pillars: the intersection with the x-axis at \SI{51}{\femto \second} corresponds to the time-of-flight (TOF) taken by the electron to travel from the large to the small pillar ($\text{TOF} = \Delta z/v_e$, with $\Delta z$ the difference in height between the pillars, and $v_e$ the speed of the electrons). The time delay of the inner band is due to the interference of radiation from the large tip and the small tip ($\Delta t_+$) and the interference of their reflecting signals ($\Delta t_-$), given by 
\begin{equation}
 \Delta t_{\pm}=\text{TOF}\pm\Delta z \sin(\theta)/c.
 \label{eq:Latcoh_eq4}
\end{equation}
The time delay of the outer bands is due to the interference of the tip radiation of one pillar with the reflected signal of the other pillar. Since the distance between the pillar and the mirror image of the other pillar ($\Delta z_{\text{ref}}=z_1+z_2$) is larger than the distance associated with the inner bands, these bands follow a different relation and the time delays associated with this interference are given by 
\begin{equation}
 \Delta t_{\pm} = \text{TOF}\pm \Delta z_{\text{ref}}\sin(\theta)/c.
\end{equation}
From the observation of the interference between radiation from both pillars, we conclude that both pillars are coherently excited. In particular, the fact that we identify a component that includes the TOF of the electron is an indication for their coherent excitation. This is the result of the fact that both plasmonic tips are excited within the evanescent field of the same electron.

\section{CL interferometry of double pillar}
\subsection{Theoretical solution}
\begin{figure}[b]
	\centering
	\includegraphics{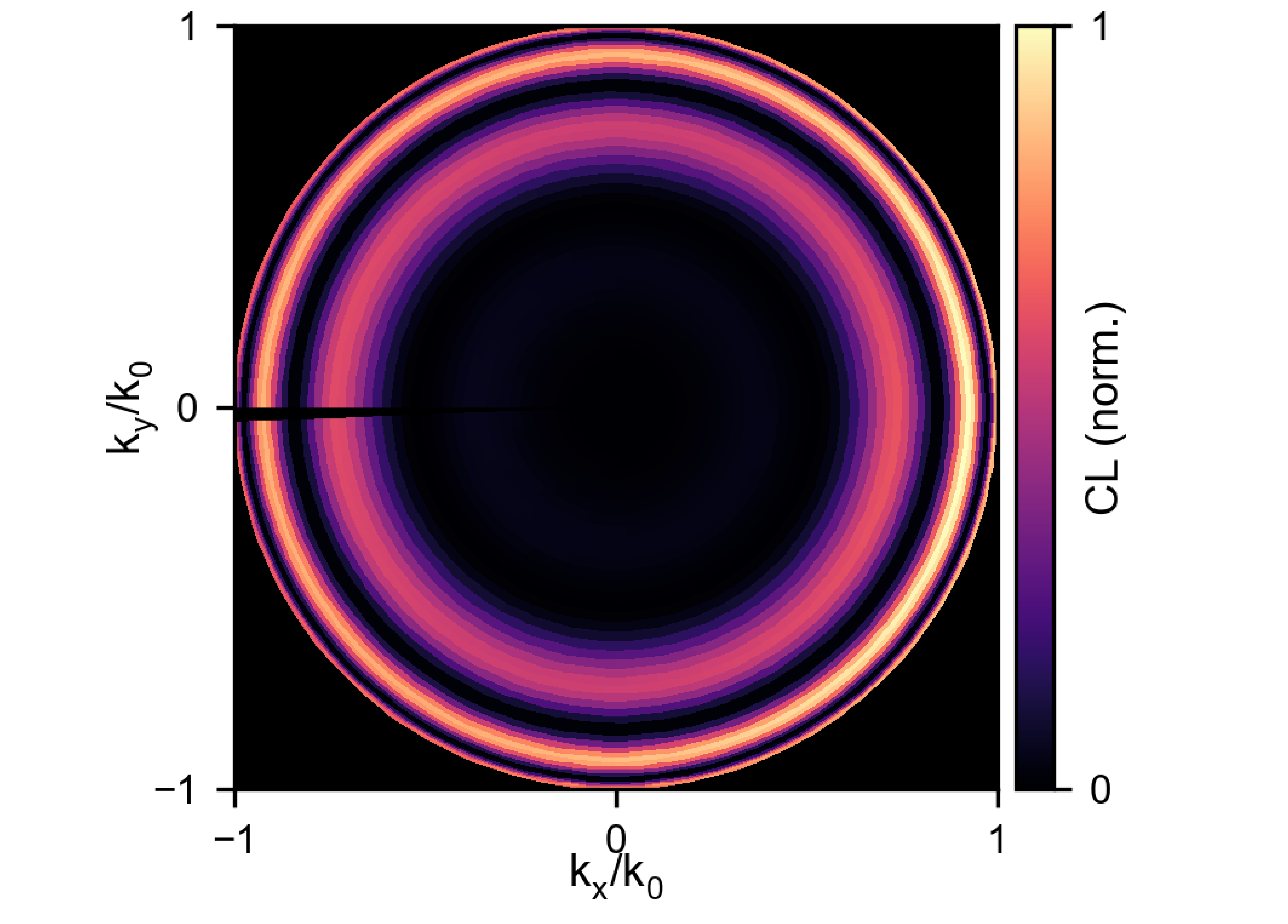}
	\caption{Calculated CL interferogram for two nanopillars with a vertical separation of \SI{1.6}{\micro\meter}, at a wavelength of \SI{500}{\nano\meter}.}
	\label{fig:Latcoh_S3}
\end{figure}

To calculate the expected CL interferogram for two separated plasmonic pillars, we use an analytical model \cite{akerboom2025}. Figure \ref{fig:Latcoh_S3} shows the resulting CL interferogram for two pillars separated by a vertical distance of \SI{1.6}{\micro \meter}, calculated at a wavelength of \SI{500}{\nano\meter}. For the calculation, the plasmonic tip was approximated as a Au particle with a radius of \SI{75}{\nano\meter} and the e-beam was positioned at \SI{10}{\nano\meter} from the particle's center.
\newpage
\subsection{Experimental data}
\begin{figure*}[t]
	\centering
	\includegraphics{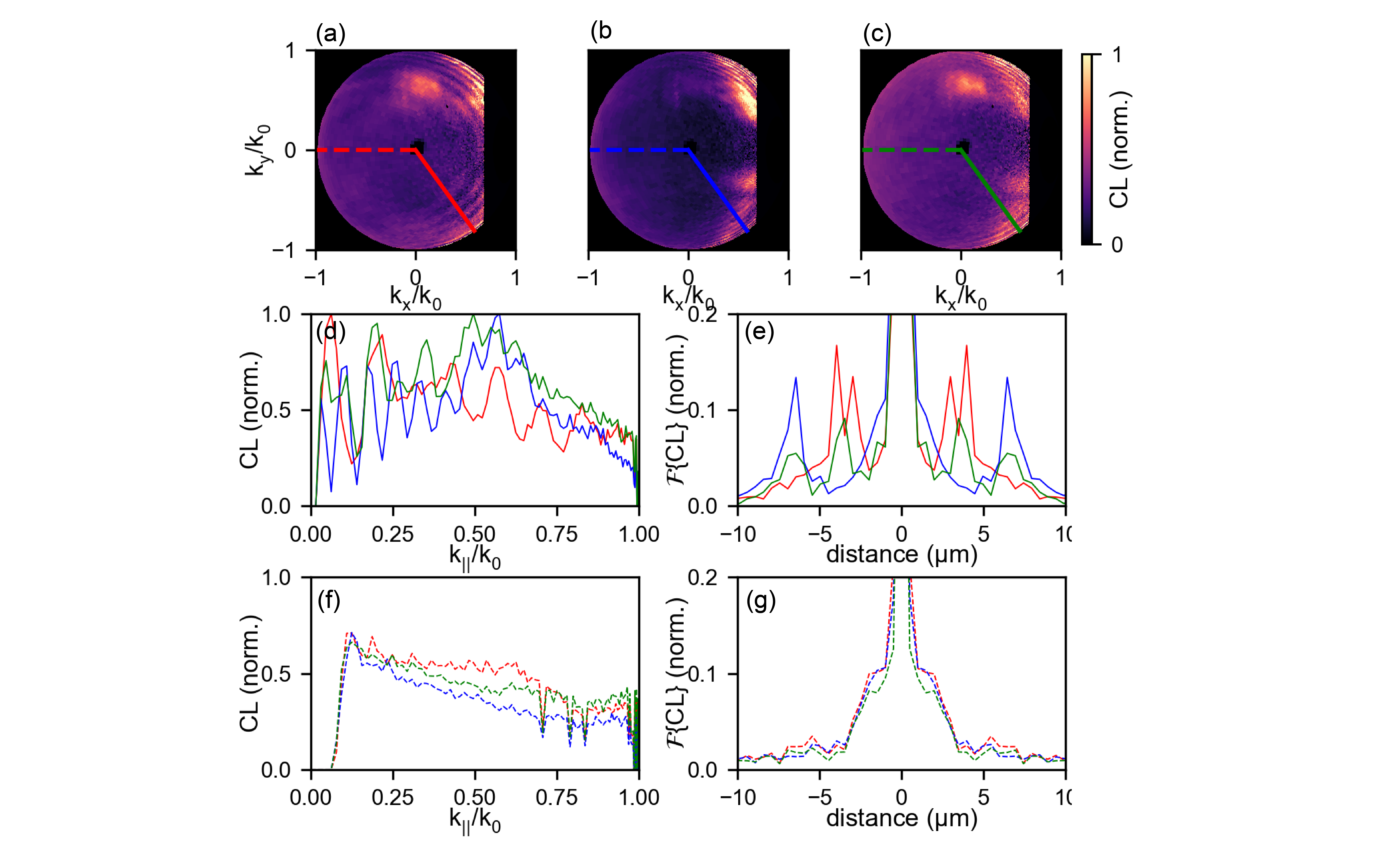}
	\caption{Excitation of a double pillar geometry with a laterally spread e-beam. (a-c) AR CL data from the double pillar geometry using a \SI{50}{}-\SI{}{\nano\meter} bandwidth filter centered around a wavelength of \SI{500}{\nano\meter}. (d,f) Line scans along the $\textbf{k}_{\parallel}$-space images and (e,g) their Fourier transform, taken from cuts along the solid lines and dashed lines in (a-c), respectively. The double pillar was excited by an e-beam passing near the smaller pillar (red) and the taller pillar (blue), or placed in the middle with an e-beam diameter of \SI{500}{\nano\meter} (green).}
	\label{fig:Latcoh_S4}
\end{figure*}

Figure \ref{fig:Latcoh_S4} shows the experimental data, similar to Fig. 3 from the main text. Figures \ref{fig:Latcoh_S4}(d) and \ref{fig:Latcoh_S4}(f) show line scans along the $\textbf{k}_{\parallel}$-space images and Figs. \ref{fig:Latcoh_S4}(e) and \ref{fig:Latcoh_S4}(g) the corresponding Fourier transform, taken from cuts along the solid lines and dashed lines in (a-c), respectively. The double pillar was excited by an e-beam passing near the smaller pillar (red) and the taller pillar (blue), or positioned between the pillars with an e-beam diameter of \SI{500}{\nano\meter} (green). When we compare Fig. \ref{fig:Latcoh_S4}(d) and Fig. \ref{fig:Latcoh_S4}(f), we observe an absence of interference fringes in the latter, while there is still light being emitted at negative $k_x$ values.

\newpage
\section{Correlation experiment to remove which-path information}
\begin{figure}[h]
	\centering
	\includegraphics{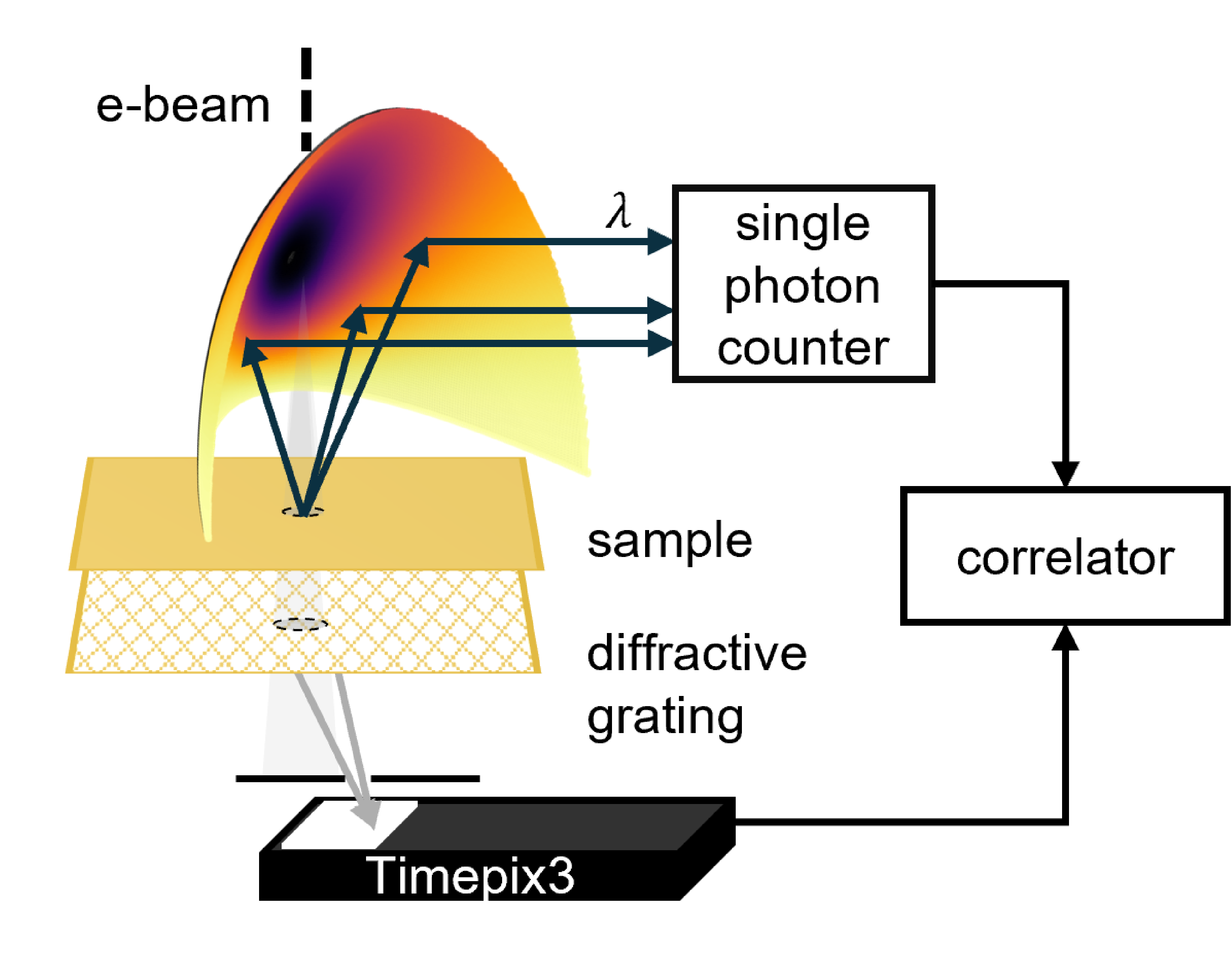}
	\caption{Schematic representation of a proposed correlation experiment to remove which-path information and retrieve CL interference. First, the e-beam interacts with a thin sample emitting CL. Subsequently, electrons are recombined using a diffractive grating, such as a bilayer of graphene, and only electrons in the overlapping diffraction regions are collected. The correlated photons are then post-selected to retrieve CL interference.}
	\label{fig:Latcoh_fig5}
\end{figure}

Figure \ref{fig:Latcoh_fig5} shows the proposed experiment to erase the electron's which-path information in order to retrieve interference in the emitted CL. Using a bilayer of graphene, electrons from different positions are mapped onto the same detector position. Using a coincidence measurement to filter CL events from only these electrons would allow the recovery of CL interference. 

\end{document}